\documentclass[aps,prl,a4paper,twocolumn,epsf,showemail]{revtex4}

\usepackage{graphicx}
\usepackage{dcolumn}
\usepackage{bm}
\usepackage{amsmath}
\usepackage{textcomp}

\begin{document}

\title{Fabrication of alignment structures for a fiber resonator by use of deep-ultraviolet lithography}

\author{Xiyuan Liu}
\email{xiyuanl@rumms.uni-mannheim.de}
\author{Karl-Heinz Brenner}
\affiliation{Lehrstuhl f\"ur Optoelektronik, Universit\"at Mannheim,
\\B6 23-29, D-68131 Mannheim, Germany.}

\author{Marco Wilzbach}
\email{wilzbach@physi.uni-heidelberg.de}
\author{Michael Schwarz}
\author{Thomas Fernholz}
\author{J\"org Schmiedmayer}
\affiliation{Physikalisches Institut der Universit\"at Heidelberg, \\
Philosophenweg 12, D-69120 Heidelberg, Germany.}

\date{Published December 2005}

\begin{abstract}
We present a novel method to mount and align an optical-fiber-based 
resonator on the flat surface of an atom chip with ultrahigh precision. 
The structures for mounting a pair of fibers, which constitute the 
fiber resonator, are produced by a spin-coated SU-8 photoresist 
technique by use of deep-UV lithography. The design and production 
of the SU-8 structures are discussed. From the measured finesses we 
calculate the coupling loss of the SU-8 structures acting as a kind 
of fiber splice to be smaller than 0.013 dB. \\

\end{abstract}

\maketitle

\section{Introduction}
Although integrated optics is a widespread and important
field in today's technology, especially in the
telecommunication sector, integrated optics is also
beginning to gain in importance in other areas, such
as atomic physics and quantum optics. In the developing
field of quantum information processing, the
manipulation of neutral atoms with so-called atom
chips provides a new, extremely promising approach \cite{folman,
folman2}. The concept relies on the capability to
control magnetically trapped atoms that hover micrometers
above a microstructured surface that carries
electric currents and charges to provide the necessary
fields. As far as we know, integrated optical access to
trapped atoms has not yet been implemented on
these atom chips.

An important step will be the on-chip detection of
single atoms that can be achieved with an optical
fiber resonator \cite{horak}. In our approach, the fibers are
mounted in a superstructure fabricated from SU-8
resist that provides positioning and alignment accurately
and easily. The fibers can easily be inserted by
hand and are automatically aligned with submicrometer
precision.

SU-8 is an epoxy-based, chemically amplified solvent-developed 
negative resist that is typically patterned
with 365-436 nm UV aligners. Its specific
properties facilitate the production of thick structures
with smooth, nearly vertical sidewalls \cite{microchem2}. Because
of the high mechanical, chemical, and thermal
stability of the polymerized SU-8, it has been used to
fabricate a wide range of microcomponents, such as
optical planar waveguides with outstanding thermal
stability and controllable numerical apertures, mechanical
parts such as microgears for engineering
applications, microfluidic systems, and microreactors
for biochemical processing \cite{ruhmen4}.

To assess the quality of the alignment structures,
we use the fiber resonator itself. Since the finesse of
the resonator strongly depends on losses introduced
by misalignment, it is a good way to measure the
coupling efficiency and alignment precision of the
SU-8 fiber splice.

\section{Fiber resonator setup}

The fiber setup is sketched in Fig. \ref{fig:liuF1}(a). We use a
Fabry-Perot-type resonator that is produced by coupling
two pieces of single-mode fiber (4.9 \textmu m mode
field diameter) with dielectric mirrors glued to the
outer ends of the fibers \cite{marco}. A small gap of a few micrometers
between the inner ends of the fibers provides
access to magnetically trapped atoms that
interact with the light field. An important property of
a resonator is its finesse which can be written as
\begin{equation}
\mathcal{F} = \frac{\delta\nu}{\Delta\nu} \approx
\frac{\pi}{\sum\limits_{i} \alpha_i},
\end{equation}
where $\alpha_i$ is the loss factor per single pass, $\delta\nu$ is the
free spectral range, and $\Delta\nu$ is the full width at halfmaximum
of the resonances. The approximation is
valid for $\mathcal{F}\gg 1$. For a more detailed description of
resonator theory we refer the reader to the textbooks
listed in Refs. \cite{demtroeder} and \cite{siegman}. With a resonator of sufficient
finesse ($\mathcal{F}>100$), the additional loss caused by light
scattering inside the gap by atoms can be used to
detect the presence of even single atoms in the gap \cite{horak}.
To obtain a high enough finesse, the fiber ends must
be aligned with submicrometer precision.

\subsection{Intrinsic losses}

In the following, all the loss mechanisms of the intact
fiber resonator are referred to as intrinsic losses, i.e.,
all the losses without introducing the gap. The losses
are basically determined by the quality of the glued
mirrors at the ends of the fiber. The loss caused by the
fiber itself is 3 dB/km as stated by the manufacturer.
With a typical length of $L=10$ cm for our resonators,
this is equivalent to a negligible loss of
0.0003 dB or 0.007 \%. In principle, the transmission
through the mirrors is determined by the properties
of the dielectric stack and can be chosen to meet
specific requirements. But the thickness of the glue
layer, the alignment precision, internal losses, and
the surface roughness limits the achievable reflectivity.
The most important limitation, which cannot be
overcome, is the spreading of the unguided light
mode in the glue layer and within the mirror itself.
This leads to a reduced coupling of the reflected light
back into the fiber.

\subsection{Losses caused by the gap}
\begin{figure}[h!]
\centering
\includegraphics[width=0.45\textwidth]{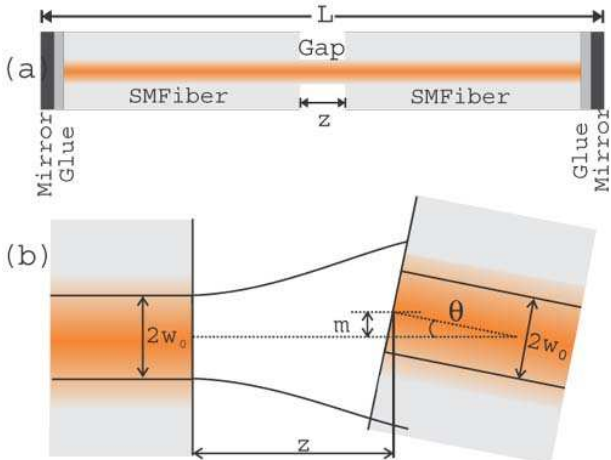}
\caption{\label{fig:liuF1} (a) Fiber resonator of length $L$, including a gap of length $z$.
Dielectric mirrors are glued to the outer ends of the resonator.
(b) Sketches of the possible misalignments at the gap. The mode of
diameter $2 w_0$ that leaves the first fiber, diverges and is partially
coupled into the second fiber, which exhibits a longitudinal displacement
$z$, a transversal displacement $m$, and an angle misalignment $\theta$.
}
\end{figure}

After cutting the resonator and introducing the gap,
the light coupling between the two pieces will be
reduced, thus introducing additional loss, which results
from light scattering at the newly introduced
surfaces and from transversal, angular, and longitudinal
misalignment. The relevant geometric parameters
are shown in Fig. \ref{fig:liuF1}(b). Rotational misalignment
converts a potentially imperfect core-cladding concentricity
into transversal misalignment \cite{saruwatari}.

For a fiber with a single-step refractive index profile, 
a Gaussian approximation \cite{ghatak} for the fundamental
mode field distribution can be used. Typically, the
Gaussian approximation deviates by less than 1\%
from the true mode field. In general, the power coupling
efficiency for two fibers can be calculated by the
overlap integral of the fiber optical field modes. The
efficiency decreases quadratically with the geometric
parameters for small deviations from perfect alignment.
We found that the crucial parameters for efficient 
light coupling are the transversal misalignment
and the angle between the optical axes of the two
fiber pieces. Because of the weak dependence of the
mode field diameter in the near field, the coupling
loss caused by longitudinal misalignment is not so
critical. One must take into account that Fresnel
backreflection at the gap surfaces leads to a coupled
system of three resonators. But the influence of longitudinal
mode symmetry on the resonator finesse
vanishes for small gap sizes.

\section{Deep-Ultraviolet lithography by use of SU-8 photoresist}
\subsection{Structure requirements}
\begin{figure}[h!]
\centering
\includegraphics[width=0.48\textwidth]{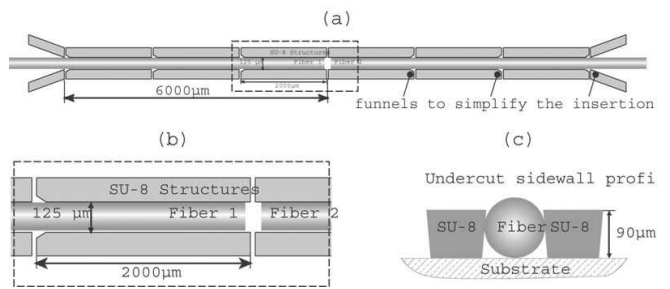}
\caption{\label{fig:liuF2} (a) Layout of the alignment structure, (b) a
magnified part (dotted rectangle), and (c) the undercut sidewall
profile.}
\end{figure}

The alignment structures for the fiber resonator must
meet some specific requirements. They must be able
to tolerate temperature changes and gradients. In
typical experiments with atoms trapped in microscopic
potentials, the currents carried by the metallic
structures lead to a local temperature increase of as
much as 100 \textcelsius. Furthermore, the structure must be
taller than the fiber radius {($r=67.5$ \textmu m)} and an
exposure in thick resist is needed. To prevent lateral
and angular misalignment, i.e., parallel and perpendicular
to the substrate plane, an undercut sidewall
profile is superior to a vertical sidewall profile. With
such a profile, the separation between the sidewalls
decreases proportional to the distance from the substrate
surface [see Fig. \ref{fig:liuF2}(c)], thus clamping the fiber.
To meet these requirements, SU-8 is highly suitable,
because of its thermal stability and outstanding lithographic
performance. The undercut sidewall profile
can be obtained by optimization of the lithographic
process steps. The optimization techniques include
fine-tuning of the exposure dose and the postexposure
bake (PEB) time.

The layout of the desired alignment structure with
fibers is shown in Fig. \ref{fig:liuF2}. This design includes funnels
to simplify the insertion of the fiber. To avoid angular
misalignment, the total length of the alignment
structure was chosen to be 6000  \textmu m, and it is divided
into several subsegments to reduce stress induced by
thermal expansion.

\subsection{Optimization of the fabrication process}
The process for the fabrication of alignment structures
includes substrate coating, soft bake, UV exposure,
PEB, and development. Each process has a
strong influence on the final structure, and there are
complex interrelations among the single process
steps. Substrate cleaning and dehydrate baking prior
to spin coating can improve SU-8 adhesion. An insufficient 
soft bake results in mask adhesion and an
uneven resist layer. On the other hand, a too long soft
bake time causes stress and cracking in the final
structures. The degree of polymerization is controlled
by both the exposure dose and the bake conditions.
All these interrelations increase the complexity of
optimization. To accelerate the optimization process,
we limited the variation of the process to parameters
that have the strongest influence on the final results.
Specific to our project, the slight undercut sidewall
profile is important and can be optimized by varying
the exposure dose and the PEB conditions. Compared
with the PEB conditions, the exposure dose has a
stronger influence on the sidewall profile, therefore
the optimization was performed by a variation of the
exposure dose.

\begin{figure}[h!]
\centering
\includegraphics[width=0.48\textwidth]{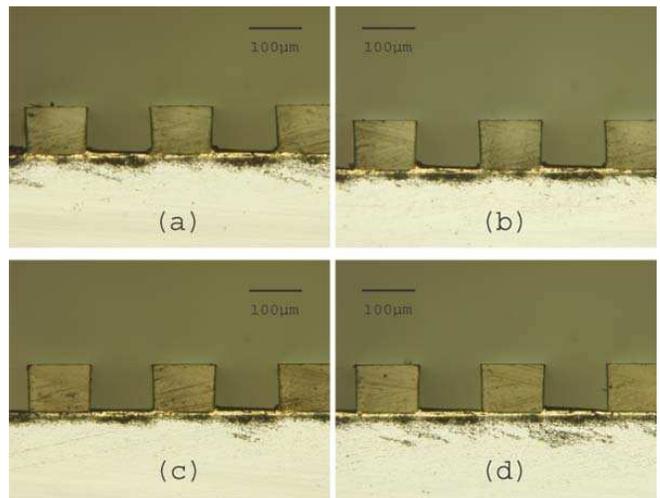}
\caption{\label{fig:liuF3} Microscope images of the cross section of SU-8
structures with exposure time increasing from (a) to (d).}
\end{figure}

The entire process is described in detail in the following.
To improve the adhesion of SU-8 films, the
gold-coated silicon substrates were cleaned in an ultrasonic
bath for 5 min at room temperature. They
were subsequently rinsed in distilled water and then
dehydrated on a hot plate at 200 \textcelsius\ for 1 h immediately
before use. After cooling to room temperature,
approximately 3 g of SU-8 50 resist was spread over
5 cm $\times$ 5 cm of the substrate around the central area.
Spin coating the resist at 500 rpm for 20 s, followed
by 2000 rpm for 20 s produced an approximately
90 \textmu m thick film. The coated film was then prebaked
on a hot plate in two steps to evaporate the solvent. In
the first step we used a temperature of 65 \textcelsius\ for
10 min. Then the temperature was ramped up to
95 \textcelsius\ for approximately 6 min and then held constant
at 95 \textcelsius\ for 2 h. After cooling to room temperature,
the substrate was exposed with the desired
pattern mask by use of a standard 365 nm UV light
source. To optimize the undercut sidewall profile, we
used a reduced exposure dose. During the PEB time,
the exposed area of the SU-8 film was selectively
polymerized. The postbake process was also performed
in two steps. The substrate was placed on the
hot plate at 65 \textcelsius\ for 1 min. This step is necessary to
avoid an image flow before the resist is slightly polymerized.
Then the substrate was immediately put on
another hot plate at 95 \textcelsius\ for 10 min. After the PEB,
the substrate was removed from the hot plate and
cooled to room temperature. Finally, the nonpolymerized
regions of the SU-8 film were removed in SU-8
developer for 12 min. To observe the sidewall profile,
the substrate was cut with a precision dicing saw.
The microscopic images of the sidewall profiles are
shown in Fig. \ref{fig:liuF3} for different exposure times. The
pictures indicate that the degree of undercut becomes
larger with smaller exposure doses. This result can be
explained by light diffraction at the mask aperture.
Because the adhesion of the resist to the substrate
decreases with lower exposure, we chose a compromise
between an acceptable undercut and a sufficient
adhesion, which corresponds to that in Fig. \ref{fig:liuF3}(b)).

\section{Results}
We determined the quality of the SU-8 fiber splice
indirectly by first measuring the finesse of an intact
resonator and then splitting and inserting it into the
structures. We recorded the transmitted light intensity
while scanning the laser over several free spectral
ranges of the fiber resonator. A model function
was fitted to the data, yielding the finesse according
to Eq. (1). The results were averaged over several
hundred runs of the experiment. The finesses of two
intact resonators were found to be $\mathcal{F}_{1}=110.4 \pm 0.3$
and $\mathcal{F}_{2}=152.8 \pm 1.1$. After cutting the resonators
and polishing the surfaces, the pieces were
introduced into the SU-8 structures. We observed the
fiber ends under a microscope and minimized the gap
sizes to touching fibers. The finesses were then measured
to be $\mathcal{F}_{1}=101.1 \pm 0.5$  and $\mathcal{F}_{2}=132.0 \pm 1.3$, thus giving an additional average loss of $\alpha=(0.29 \pm 0.04) \%$ or ($0.013 \pm 0.002$) dB. Neglecting
other additional losses, this corresponds to a pure
lateral misalignment of $m=150$ nm or a pure angular
misalignment of $\theta=6.3\times 10^{-3}$ rad $\approx
0.36^{\circ}$. To
test thermal stability, we varied the temperature of
the substrate between 20 and 70 \textcelsius. The finesse of
the inserted fiber resonator showed no change during
heating.

\section{Conclusion}

In summary, we have demonstrated a method for
aligning fibers on a flat surface by using SU-8 superstructures.
The aligned fibers represent a Fabry-Perot-type 
resonator for atomic physics to detect
atoms. We have investigated the different loss mechanisms
for this type of fiber resonator. We then introduced
the layout of the SU-8 alignment structures,
which enables easy positioning and alignment, and
because of the undercut sidewall profile, also offers a
method of fixing the fiber position. To achieve this
structure, we optimized the lithographic process.
Furthermore we demonstrated a technique for quantifying
the losses that are due to misalignment with
the help of the fiber resonator itself. The finesse measurement
indicated that the SU-8 superstructures
are of superior quality.

\section{Acknowledgements}
We thank S. Groth for supplying gold-coated silicon substrates.
This research was partly supported by European Union contract 
IST-2001-38863 (Atom Chip Quantum Processor collaboration) and 
the Landesstiftung Baden/W\"urttemberg
Forschungsprogramm Quanteninformationsverarbeitung.


\begin{thebibliography}{99}
\bibitem{marco} M. Wilzbach, Physikalisches Institut der Universit\"at Heidelberg,
Philosophenweg 12, D-69120 Heidelberg, Germany, is preparing a
manuscript to be called ``Building micro-cavities with optical
fibers for single atom detection.''
\end{thebibliography}
\end{document}